\documentstyle[aps,prb,multicol,eqsecnum]{revtex}
\begin{document}
\draft

\title{Random Exchange Quantum Heisenberg Chains}

\author{A.\ Furusaki,\cite{AF} M.\ Sigrist, E.\ Westerberg, P.\ A.\ Lee}
\address{Department of Physics, Massachusetts Institute of Technology,
Cambridge, MA 02139}

\author{K.\ B.\ Tanaka\cite{KT} and N.\ Nagaosa}
\address{Department of Applied Physics, University of Tokyo, Hongo,
Bunkyo-ku, Tokyo 113, Japan}

\date{\today}

\maketitle

\begin{abstract}
The one-dimensional quantum Heisenberg model with random $\pm J$ bonds
is studied for $S=\frac{1}{2}$ and $S=1$.
The specific heat and the zero-field susceptibility are calculated
by using high-temperature series expansions and quantum transfer
matrix method.
The susceptibility shows a Curie-like temperature dependence at low
temperatures as well as at high temperatures.
The numerical results for the specific heat suggest that there are
anomalously many low-lying excitations.
The qualitative nature of these excitations is discussed
based on the exact diagonalization of finite size systems.
\end{abstract}


\begin{multicols}{2}

\section{Introduction}
One-dimensional (1D) quantum spin systems have attracted much interest
for many years.
They provide a very useful test ground for our understanding
of quantum systems with many degrees of freedom and have led to the
development of various powerful methods of theoretical physics.
Quantum spin systems are also of interest in their own right for the large
variety of their ground states and properties of excitations.
Over the years the search for real spin chains has led to
the synthesis of various quasi-1D systems.
In general we encounter defects and disorder in real systems,
which may destroy much of the properties expected for uniform spin chains.
Although at first sight unwanted, these defects have initiated a great
deal of studies on disordered chains of classical as well as quantum
spins.

To our knowledge the first system investigated in this respect
originates from the quasi-1D organic charge transfer compounds
of tetracyanoquinodimethanide (TCNQ) which was described by
the nearest-neighbor Heisenberg model,
\begin{equation}
{\cal H}=\sum^{L-1}_{i=0}J_i{\bf S}_i\cdot{\bf S}_{i+1},
\label{H}
\end{equation}
where the interaction is antiferromagnetic but random
in strength with a continuous broad probability
distribution, $P(J_i)$.\cite{Bulaevskii}
This type of model was investigated by Hirsch and Jos\'e\cite{Hirsch}
by means of renormalization group (RG) methods based on
a Kadanoff block spin decimentation scheme.
Ma and coworkers,\cite{Ma} used a different real space RG method
where the strongest bonds were successively integrated
out in order to reach an effective Hamiltonian with a fixed point
of the distribution $P(J_i)$.
Fisher\cite{Fisher} recently revisited this RG method and extended it
also to the XXZ Heisenberg model.
The result of this analysis is that the ground state is a random singlet
phase\cite{Bhatt} where each spin is paired into a singlet with another
spin which may be located far away in the chain.

Another approach for the disordered spin chains is the mapping of
the spins into fermions via a Jordan-Wigner transformation,
which in turn are treated by a bosonization scheme.
This approach was used by Nagaosa\cite{Nagaosa} to study the effect of
a random magnetic field.
He derived scaling relations for spin-spin correlation functions
analytically and confirmed them by numerical calculations.
Assuming weak disorder, Doty and Fisher\cite{Doty} then
analyzed the phase diagram and thermodynamic properties of XXZ spin
chains with various type of disorder by applying a perturbative RG method
to the bosonized Hamiltonian.
These results have been confirmed to some extent by extensive numerical
calculations for finite systems by Haas {\it et al.}\cite{Haas} and
by Runge and Zimanyi.\cite{Runge}

A new class of random spin chains has been found in the compound
${\rm Sr}_3{\rm Cu}{\rm Pt O}_6$ where the Cu-site alternates
with the Pt-site along chains.\cite{Wilkinson}
The Cu has a spin-$\frac{1}{2}$ degree of freedom while Pt has none.
The interaction between the Cu spins is antiferromagnetic.
Measurments of the uniform susceptibility show that this
system is well described by a 1D antiferromagnetic (AF) Heisenberg model.
The Pt atoms can be replaced by Ir, which carries a spin
$\frac{1}{2}$ like Cu.
In the chains the spin on an Ir atom
couples {\it ferromagnetically} to its neighboring Cu spins.
As a result the number of spins has doubled and the interaction
is ferromagnetic.
This spin chain is described well by the 1D ferromagnetic (FM)
Heisenberg model.
It is also possible to produce an alloy by replacing a fraction $1-x$
of Pt by Ir:
${\rm Sr}_3{\rm Cu}{\rm Pt}_x{\rm Ir}_{1-x}{\rm O}_6$, where
$x$ can be chosen at will.\cite{Nguyen}
This yields a system with random FM and AF interaction of fixed
strength.

Motivated by these experiments we consider the Heisenberg model,
Eq.~(\ref{H}), where the probability distribution of $J_i$ has the
following form,
\begin{equation}
P(J_i) = p\,\delta(J_i+J) + (1-p) \delta (J_i - J).
\label{P(J_i)}
\end{equation}
The coupling $J_i$'s have the same strength, but are
random in sign, with a probability of $p$ ($0<p<1$) to be ferromagnetic
and $1-p$ to be antiferromagnetic.
We note that this model does not exactly realize the conditions found
in the alloys described above.
In the real system the strengths of the FM and the AF bonds are in
general different.
Additionally, there is the correlation (neglected in our model) that
FM bonds appear always in even numbers.
Nevertheless, we believe that the characteristics of our model are very
similar to the ones of the real alloy and that the properties
which determines the physics is the discrete randomness with FM and
AF bonds.

Obviously the analytical treatments mentioned above are not immediately
useful for our model.
The scheme by Ma and coworkers cannot be simply applied here,
because there are no strongest bonds which would allow the
decimation procedure.
Furthermore, the disorder is not weak in our case, and thus perturbative
treatments would not lead to sensible results.
Instead we apply various types of numerical methods, such as
high-temperature expansion (HTE), transfer matrix method (TMM), and
exact diagonalization (ED) of small systems.
The first two provide accurate results over a wide range of temperature
($k_BT>0.1J$) and the last one can be used to reach a qualitative
understanding of certain aspects of the low-temperature regime.

Before going into technical details we will briefly outline
the picture which emerges from our analysis.
The main result is that this quantum spin system exhibits {\em three}
distinct regimes, high-, intermediate- and low-temperature regime.
It will become clear that this property originates from the fact
that the spin chain consists of alternating FM and AF segments of
variable length.
At very high temperatures ($k_B T\gg J$) the interaction among the
spins does not play any role and the spins behave independently.
Hence, the randomness of the exchange does not play any role in the
physical properties.
In particular, the uniform susceptibility follows the ordinary Curie law,
$\chi\propto 1/T$.
When the temperature is lowered, the spins gradually start to correlate.
In the AF segments the spins form collectively the smallest possible total
spin (either a singlet, $S=0$, or a doublet, $S=\frac{1}{2}$),
while the FM segments build up a large spin degree of freedom by
aligning spins parallel to each other.
Due to the quantum nature of the spins the interaction among these new
effective degrees of freedom is rather weak.
Therefore, in an intermediate temperature range ($k_BT\lesssim J$),
they behave like independent spins and lead to a new Curie behavior of
the susceptibility, however with a different (effective) Curie constant.
The excitations relevant in this temperature regime are dominated by the
discrete spectrum of ``spin wave'' modes of each segment.
These intrasegment modes are localized because of the mismatch of
momentum and energy between the modes in different segments.
Thus these modes cannot be transfered easily between different segments.
The thermodynamics of this regime is effectively described by an ensemble
of decoupled FM and AF segments of finite length.
Only at very low temperatures ($k_B T\ll J$) the interactions
between spins belonging to different segments become important.
Within our treatments this low-temperature regime is least understood
at present.
However, our results suggest that a considerable fraction of density of
states is located at low energies ($\omega\ll J$).
It is characterized by an effective model of spins with various sizes
which are coupled by either ferromagnetic or antiferromagnetic
interaction of widely distributed strengths.

The crossover between the regimes manifests itself in
peak-like structures of the specific heat.
These structures are located where spins begin to correlate.
While the uniform FM and AF systems have only one such (broad) peak
(corresponding to the correlation of the $S=\frac{1}{2}$ spins),
we expect that two structures appear in the random system,
a peak at temperature where the original $S=\frac{1}{2}$ spins start to
correlate ($k_B T\sim J$) within each segment,
and, possibly, a shoulder where the effective spin
degrees of freedom in each segment start to correlate.
The location and size of the second structure will depend on $p$.

The existence of three regimes is in contrast to the situation
in the analogous classical case, where only two regimes exist.
The reason for this difference lies in the following properties
of the classical spin system.
In the case of classical spins the bond disorder of the type of
Eq.~(\ref{P(J_i)}) is irrelevant for the thermodynamic properties.
A simple transformation of the spin variables (changing the sign of
some of the spins) yields a uniform FM spin chain without affecting
the energy spectrum.
Consequently, the specific heat would not depend on the disorder and
occurrence of FM and AF segments does not have any direct implications
here.
However, bond disorder affects the spin-spin correlation and
the susceptibility.
The susceptibility $\chi$ shows a clear signature of a crossover from
a high-temperature regime to a new regime at lower temperature similar
to that seen for the quantum spin system.
Despite this qualitative similarity the nature of the crossover is
quite different.
The classical spins do not separate into segments, but they correlate
within a length $\tilde{\xi}$, the correlation length of the
uniform system, which is independent of disorder.
As this length grows with decreasing temperature, the spins in a cluster
of length $\tilde{\xi}$ act together as one large effective spin whose
size scales as $S_{\rm eff}\propto\tilde{\xi}^{1/2}$ for $\tilde{\xi}$
much larger than the lattice constant.
(This follows from the fact that the directions of the spins are random
depending on the signs of the bonds so that a random walk picture applies).
These effective spins behave independently and the susceptibility per
lattice site shows the Curie behavior in the low-temperature limit:
\begin{equation}
\chi\propto\frac{S^2_{\rm eff}}{T\tilde\xi(T)}\propto\frac{1}{T}.
\label{chi-classical}
\end{equation}
It is obvious that there are no further correlation
effects beyond this, so that this second Curie regime represents
indeed the low-temperature regime reaching down to $ T=0K $.

This paper is organized as follows.
In Sec.~II we consider technical aspects of the HTE and the TMM,
in particular, the problem of extrapolation from the high-temperature
limit to low temperatures.
Readers who are not interested in technical details can skip this section.
In Sec.\ III we present the numerical results of the HTE and the TMM.
Section IV is devoted to the results of the ED of small clusters.
We demonstrate the weakness of the coupling between the effective spins
formed in the segments.
Finally, we discuss our results in Sec.\ V.

\section{Methods of numerical analysis}
In this section we describe some technical details of the HTE and the TMM.
The former allows for an exact ensemble average over the random exchange
for arbitrary $p$.
We use the latter method only for $p=0.5$ to check the validity of the
extrapolation scheme used in the HTE.
It turns out that both methods give good quantitative results over
a wide temperature range $k_BT\gtrsim0.1J$.

\subsection{High-temperature series}
The high-temperature expansion of spin systems can be implemented
as a cluster expansion algorithm in a straightforward way.\cite{Gelfand}
In 1D this algorithm is greatly simplified.

Consider a cluster of length $n$ which is described by Hamiltonian,
\begin{equation}
{\cal H}_n=\sum^{n-1}_{i=1}J_i{\bf S}_i\cdot{\bf S}_{i+1},
\label{H_n}
\end{equation}
and assume that $m$ of $(n-1)$ exchange couplings, $J_i$, are ferromagnetic
and the others are antiferromagnetic.
There are ${}_nC_m=n!/m!(n-m)!$ different configurations of this type.
For each configuration of couplings $\{J_i\}$ we calculate thermal average
of a physical quantity in powers of the inverse temperature $\beta$:
\begin{eqnarray}
o_{m,n}(\beta;\{J_i\})&=&
\frac{{\rm Tr}({\cal O}_n e^{-\beta {\cal H}_n})}
     {{\rm Tr}(e^{-\beta{\cal H}_n})}\cr
&=&
\frac{{\rm Tr}[{\cal O}_n\sum_l(-\beta{\cal H}_n)^l/l!]}
     {{\rm Tr}[\sum_l(-\beta{\cal H}_n)^l/l!]}\cr
&=&
\sum_l\tilde{o}_{l,m,n}(\{J_i\})K^l,
\label{omn}
\end{eqnarray}
where $K=\frac{1}{2}\beta JS$.
The thermodynamic quantities we consider are the internal energy $u$
and the uniform susceptibility $\chi$, for which the operator
${\cal O}_n$ is ${\cal O}_n={\cal H}_n$ and
${\cal O}_n=\beta\mu^2(\sum^n_{i=1}S^z_i)^2$ ($\mu$: Bohr magneton).
The coefficients $\tilde{o}_{l,m,n}(\{J_i\})$ have to be evaluated
for all configurations $\{J_i\}$.
However, some operations (such as reflection in the center of the cluster
and change of the sign of $J_i$) allow us to reduce the number of
configurations we need to calculate.

We then average over all the configurations for given $p$:
\begin{eqnarray}
o_n(K;p)&=&
\sum_lK^l\sum^n_{m=0}p^m(1-p)^{n-m}\cr
&&\qquad\times\frac{1}{{}_n C_m}\sum_{\{J_i\}}\tilde{o}_{l,m,n}
(\{J_i\}).
\label{on}
\end{eqnarray}
In the next step we recursively subtract the contributions of all
subclusters from $o_n(K;p)$:
\begin{equation}
o'_n(K;p) = o_n(K;p)-\sum^{n-1}_{k=1}(n-k+1)o'_k(K;p).
\label{o'n}
\end{equation}
Note that this subtraction is only possible after the average has been
taken.
Summing up the series $o'_n(K;p)$ yields the final series
\begin{eqnarray}
o(K;p)&=&\sum^N_{n=2}o'_n(K;p)\cr
&=&\sum_l\sum_mo(l,m)K^lp^m.
\label{o}
\end{eqnarray}
For 1D systems it turns out that the high-temperature series given by
(\ref{o}) is correct up to the order $K^{2N}$ for $u$, but only up to
$K^N$ for $\chi$.\cite{Rushbrooke}
The largest system size we considered is $N=11$ ($N=8$) for the
spin-$\frac{1}{2}$ (1) case.
In Tables \ref{table:u(l,m)s=1/2}$\sim$\ref{table:chi(l,m)s=1} we give
lists of the expansion coefficients $u(l,m)$ and $\chi(l,m)$.
For the spin-$\frac{1}{2}$ AF chain ($p=0$) our result agrees with
the earlier calculation by Baker {\it et al.}\cite{Baker}

\subsection{Analysis of the series}
The high-temperature series of finite length by itself does not
give a reliable result for $k_BT\lesssim J$.
There are, however, various extrapolation schemes which can provide
a very good approximation down to rather low temperatures.\cite{Guttmann}
Since no finite-temperature transition is possible for our 1D spin systems,
one should, in principle, be able to extrapolate the high-temperature
series down to zero temperature.

For our analysis we use the Pad\'e approximation method.\cite{Guttmann}
Although at first sight it seems straightforward to apply it to
the series of the internal energy $u(K;p)$, we encounter the following
severe problem, which comes from the very fact that the series could be
extrapolated down to zero temperature or $K\to\infty$.
{}From the relation $u(K;p)=-u(-K;1-p)$, we see that the series
$u(K;x)$ for the spin chain with $p=x$ also describes the spin chain
with $p=1-x$ on the negative real axis.
In particular, $u(-\infty;p)$ is a modulus of the ground state energy of
the latter system while $u(\infty;p)$ is the ground state energy of the
former.
That is, the series must have different limits for $K\to\pm\infty$.
It is impossible to deal with this feature in the standard Pad\'e
approximation scheme.

To circumvent this difficulty, we use the following method.
We first note that $u(K;p)$ is a monotonic function of $K$ along
the real axis.
Thus we can invert the series $u(K;p)$ to $K(u;p)$, a series in powers
of $u/J$.\cite{Honda}
The function $K(u;p)$ vanishes at $u=0$ and diverges both at
$u=u(\infty;p)$ and at $u=-u(\infty;1-p)$.
Thus the temperature range $-\infty<K<\infty$ is mapped to the region
$u(\infty,p)\le u\le-u(-\infty;1-p)$.

The inverted series $K(u;p)$ can now be analyzed by means of Pad\'e
approximants\cite{differential} because the two limits $K\to\pm\infty$
correspond to two finite points.
If we assume that near zero temperature the internal energy behaves as
$u(K;p)-u(\infty;p)\propto T^{1+\alpha}$, or equivalently
\begin{equation}
K\propto[u-u(\infty,p)]^{-\frac{1}{1+\alpha}},
\label{K}
\end{equation}
then the ground state energy $u(\infty;p)$ should show up in
$\frac{dK}{du}/K$ as a pole on the real axis.
Thus we can estimate the ground state energy and the exponent $\alpha$
from Pad\'e approximants of $\frac{dK}{du}/K$.
The ground state energy we estimated from these Dlog Pad\'e approximants
is given in Table \ref{table:energy}.
We expect that the estimated numbers have errors of order 1\% except
$p=1$, for which the Pad\'e approximants have disturbing poles
close to the physical one corresponding to the ground state energy.

It turns out, however, that we cannot obtain reliable estimates for
the exponent $\alpha$ in this way; the estimated values seem to be
too large.
For example, for $p=0$ Pad\'e approximants give $\alpha\approx1.85$,
which differs significantly from the exact value $\alpha=1$.
This failure in estimating $\alpha$ is possibly related to
the fact that the singularities at $u=u(\infty;p)$ and $-u(\infty;1-p)$
in the Dlog Pad\'e approximants are not simple poles.
In fact, the approximants have many poles on the real axis,
suggesting that there are two cuts along the real axis:
$u<u(\infty;p)$ and $u>-u(\infty;1-p)$.
We suspect that the exponent is more sensitive to the disturbing
cuts than the location of poles (ground state energy) is.

Now we are ready to explain how the specific heat and the susceptibility
are calculated from the high-temperature series.
By integrating the Pad\'e approximants for $\frac{dK}{du}/K$,
we first obtain $K(u;p)$.
The specific heat is then given by
\begin{equation}
C\biglb(u(T)\bigrb)=-[K(u)]^2\frac{du}{dK}.
\label{C(u(T))}
\end{equation}
Due to the too large $\alpha$, the specific heat calculated from
Eq.\ (\ref{C(u(T))}) is too small at very low temperatures.
As we will see in the next section, however, it still gives reasonable
results for $k_BT\gtrsim0.1J$.

The susceptibility is calculated from the high-temperature series
$\chi(K;p)$ in the following way.
We first transform the series of $T\chi$, which is a power series
of $K$, to a new series in powers of $u$ by substituting the inverted
series $K(u)$ into $T\chi(K)$.
We then make a Pad\'e approximant for $\frac{d}{du}T\chi(u)/T\chi(u)$
and integrate it back to get an approximant for $T\chi(u)$, which
we denote $P(u)$.
The susceptibility is finally calculated from
$\chi(T)=\beta P\biglb(u(T)\bigrb)$.
The results obtained in this way are shown in the next section.

\subsection{Quantum transfer matrix method}
In addition to the high-temperature series expansion, we use the quantum
transfer matrix method, which is also a numerical method valid
at high temperatures, to check the validity of the extrapolation
scheme used in analyzing the high-temperature series.
The basic idea of the method is to calculate the partition function
by using the Trotter breakup,
\begin{equation}
\exp(-\beta{\cal H})\approx
[\exp(-\beta{\cal H}_{\rm e}/N_T)
 \exp(-\beta{\cal H}_{\rm o}/N_T)]^{N_T},
\end{equation}
where ${\cal H}={\cal H}_{\rm e}+{\cal H}_{\rm o}$
with a checkerboard decomposition.
Then the system can be viewed as a two-dimensional classical system,
and the partition function is calculated by multiplying a transfer matrix.
The details of the method can be found in the literatures.\cite{Suzuki1}
This method has the following advantages:
1) The CPU time grows only linearly with the system size $L$ so that
we can study large systems.
The largest system we studied has 400 sites, for which we can expect
self-averaging.
2) Although the Trotter number $N_T$ is limited by the memory size of
the computer [$N_T=10$ (6) for $S=\frac{1}{2}$ (1) in our calculation],
it is possible to extrapolate to the $N_T\to\infty$ limit,
from the free energy $F_{N_T}$ calculated for a finite $N_T$,
by\cite{Suzuki2}
\begin{equation}
F_{N_T}=F_\infty+\frac{a_1}{N^2_T}+\frac{a_2}{N^4_T}+\ldots.
\end{equation}
For example, for the spin-$\frac{1}{2}$ case we use $F_{10}$, $F_9$, and
$F_8$ to determine $F_\infty$, $a_1$, and $a_2$.
This method works well down to $k_BT\sim 0.1J$.
3) In contrast to the quantum Monte Carlo method, there is no statistical
error in the transfer matrix method.
Thus, one can calculate the specific heat and spin susceptibility
by numerical differentiation with respect to the temperature and
a magnetic field.

\section{Thermodynamic Properties}
In this section we discuss the results for two measurable
quantities, the specific heat and the susceptibility,
for the spin-$\frac{1}{2}$ and spin-1 chains.

\subsection{Specific heat}
As described in Sec.\ IIB, the specific heat is calculated from the
high-temperature series by Eq.\ (\ref{C(u(T))}).
We show the results for $p=0$, 0.25, 0.5, 0.75, and 1 in
Figs.\ \ref{fig:c-s=1/2} and \ref{fig:c-s=1}.
Note that the results for the spin-$\frac{1}{2}$ AF chain ($p=0$) and
for the FM chain ($p=1$) are due to two-point Pad\'e approximants
which we impose to have poles at the known ground state energy
with the correct exponent $\alpha$ ($\alpha=1$ for $p=0$ and
$\alpha=0.5$ for $p=1$).
The quality of our approximants is demonstrated by the comparison with
the data obtained from finite size calculations by Bl\"ote\cite{Blote}
for $p=0,1$ and by the data we obtained using the TMM for $p=0.5$.
We see that our Pad\'e approximants are quite reliable
at least above $k_BT\sim 0.1J$ for $S=\frac{1}{2}$ and
$k_BT\sim 0.2J$ for $S=1$.

However, we find that our approximants for $0<p<1$ cannot be valid
in the whole temperature range;
they do not satisfy the sum rule,
\begin{equation}
S_\infty\equiv\int^\infty_0\frac{C(T)}{T}dT=k_B\ln(2S+1).
\label{sum}
\end{equation}
This can be clearly seen from Figs.\ \ref{fig:c/T-1/2} and
\ref{fig:c/T-1}, where we show $C(T)/T$ as a function of $T$.
Some entropy is missing if we naively extrapolate our high-temperature
series down to zero temperature.
Rough estimates give the following values for the missing entropy:
in the spin-$\frac{1}{2}$ (spin-1) system $\Delta S/S_\infty=$12\%,
24\%, and 36\% (14\%, 28\%, and 32\%) for $p=$0.25, 0.5, and 0.75,
respectively.
Since our results are reliable above $k_BT\sim(0.1J-0.2J)$, we expect
that the missing entropy is ``hidden'' at lower
temperatures.\cite{Furusaki,Reed}
A considerable fraction of the density of states would be located at
low energies.
We discuss in Sec.\ IV the nature of the low-lying excitations
which are responsible for the missing entropy.

\subsection{Susceptibility}
We plot the susceptibility calculated from the HTE in
Figs.\ \ref{fig:chi-1/2} and \ref{fig:chi-1}.
For $p=0.5$ we also plot the data obtained from the TMM.
The results obtained from the two different methods agree quite well
so that we can trust our data down to low temperature $\sim0.1J$.
We note that the convergence of the data of the susceptibility in
the TMM is much better than that of the specific heat.

{}From Figs.\ \ref{fig:chi-1/2} and \ref{fig:chi-1} we see that
the susceptibility obeys the Curie law at low temperatures
as well as at high temperatures.
There is a crossover at $k_BT\sim J$, and the Curie constant at lower
temperature depends on $p$.
Interestingly enough, this feature is qualitatively the same as that
of the classical Heisenberg spin chain:
\begin{equation}
\chi_{\rm cl}(T)=
\frac{\mu^2 S^2}{3 k_B T}
\frac{1+\overline{v}(JS^2/k_BT)}{1-\overline{v}(JS^2/k_BT)}
\label{classic}
\end{equation}
with
\begin{equation}
\overline{v}(x)=(2p-1)\left({\rm coth}(2x)-\frac{1}{2x}\right).
\end{equation}
Equation (\ref{classic}) is obtained\cite{Tonegawa,Fahnle,Furusaki}
by using Fisher's method.\cite{MEFisher}
For $0<p<1$ the susceptibility shows a Curie-like behavior for
both $T\to\infty$ and $T\to0$:
\begin{equation}
\chi_{\rm cl}(T)=\frac{\mu^2S^2}{3k_BT}\qquad{\rm for}~T\to\infty
\end{equation}
and
\begin{equation}
\chi_{\rm cl}(T)=
\frac{\mu^2S^2}{3k_BT}\frac{p}{1-p}\qquad{\rm for}~T\to0.
\end{equation}
Obviously there is a crossover between the two regimes at $k_BT\sim J$
where the Curie constant changes its high- to its low-temperature value.
Note that at $p=0.5$ the susceptibility of the spin chain is
the same as that of a free spin.
It is also worth noting that for the uniform system the low-temperature
susceptibility is not Curie-like, but goes to a constant for $p=0$ and
diverges quadratically for $p=1$, which is qualitatively the same
behavior as the spin-$\frac{1}{2}$ chains.\cite{Takahashi}

Besides the qualitative similarities, there are clear differences
between the quantum and the classical spin chains.
As discussed in Sec.~I, the qualitative nature of the crossover is
completely different.
For the classical spin chains the existence of the length scale
$\tilde\xi$ is essential, whereas the segmantation is crucial for
the quantum spin chains, as will be demonstrated in the next subsection.
A quantum effect is clearly visible for $p=0.5$ where, in contrast to
the classical case, we find a crossover at $k_BT\sim J$ between the
high-temperature Curie law,
\begin{equation}
\chi=\frac{\mu^2S(S+1)}{3k_BT},
\end{equation}
and a low-temperature Curie-like behavior.
For larger spins, however, the change becomes less pronounced
as we can see in Fig.~\ref{fig:chi-1}.

Unfortunately, we cannot simply claim from our numerical result that
the susceptibility diverges as $1/T$ for $T\to0$ like in the classical
system.
Although we are confident about the accuracy of our data above
$k_BT\sim0.1J$ from the comparison between the HTE and the TMM,
we cannot assume that our extrapolation scheme is valid down to
zero temperature, which is certainly not the case for the specific heat.
To determine the $T$-dependence of the susceptibility at $T\to0$,
we need to develop a completely different approach which can deal with
low-energy excitations directly.

\subsection{Discussion of the results}

In the introduction we mentioned that the key to the understanding
of our random quantum spin system lies in the property that the spin
chain consists of a sequence of alternating AF and FM segments
of various lengths.
On an intermediate energy scale they behave like independent finite size
systems.
Thus, for low enough temperature the spins within each segment correlate
into the ground state as if the segments were decoupled.
There are small thermal fluctuations to the excited states which are
separated by a finite energy gap because of the finite length of each
segment.
The ground state of the AF segments is a spin singlet or doublet
(triplet) for an odd or an even number of bonds in the spin-$\frac{1}{2}$
(spin-1) system.
In the FM segment the ground state is characterized by the formation of
the largest possible total spin, where the boundary spins should not be
included, i.e., the total spin of a segment with $\ell$ bonds is
$S_{\rm tot}=S(\ell-1)$.
It is important to realize that the boundary spins ``shared'' by
adjacent FM and AF segments do not contribute to $S_{\rm tot}$ of the
FM segment, but always have to be counted to the AF segment
for the formation of its ground state.
This fact is confirmed by analytic arguments and finite size calculation,
which will be dicussed in the next section.

For short segments the intrasegment excitation energies are of the
order of $J$.
For segments of larger length $\ell$ the lowest excitation energy
can be estimated based on a spin wave picture in a finite chain,
which scales as
\begin{equation}
\Delta E \propto \left\{
\begin{array}{ll}
\displaystyle
\frac{J}{\ell^2} \hskip 0.8 cm & \mbox{in FM segments,} \\ & \\
\displaystyle
\frac{J}{\ell} & \mbox{in AF segments.}
\end{array} \right.
\end{equation}
The intrasegment spin-wave-like modes are localized within each
segment, because the excitation spectra in neighboring segments
are different.
Therefore we expect that also the coupling between the effective spin
degrees of freedom (formed in the ground state of each segment) is
very weak.
On an intermediate energy scale ($\lesssim\Delta E$) they may be
considered as independent spins.
Only for much lower energies correlation would develop among them as
will be demonstrated in the next section.

These properties allow us to interpret some of our results of the
HTE and the TMM.
Let us first consider the specific heat.
As shown in Figs.~\ref{fig:c-s=1/2} and \ref{fig:c-s=1}, the specific heat
of disordered spin chains ($0<p<1$) drops more rapidly at low temperatures
than that of uniform chains ($p=0,1$).
As a consequence the entropy we obtained numerically does not satisfy the
sum rule, $S_{\infty} = k_B\ln(2S+1)$ per site.
{}From our discussion it becomes clear that the effective spin degrees of
freedom do not correlate in the temperature range covered by the HTE.
Therefore their contribution to the entropy is not visible in the
extrapolated specific heat.
Instead they would appear at lower temperature which is beyond reach
of our methods.
A simple estimate of the missing entropy can be given by regarding
all the segements to be independent.
In this picture a FM segment contributes the entropy
$\sigma_F=k_B\ln[2S(\ell-1)+1]$ and an AF segment
$\sigma_A=0$ ($\ell$ odd) and $k_B\ln(2S+1)$ ($\ell$ even).
The total entropy contribution per site is
\begin{equation}
\Delta \sigma =\frac{\sum_{i} \sigma_i}{N},
\label{eq:entro1}
\end{equation}
where the sum is over all the segments and $\sigma_i$ is the entropy of
segment $i$ as discussed above.
We note that $N=\sum_i n_i$.
Splitting the sums into sums over AF segments and FM segments
and using the fact that the number of AF segments equals the number of
FM segments, we obtain
\begin{equation}
\Delta\sigma=
\frac{\langle\sigma_A\rangle+\langle\sigma_F\rangle}
     {\langle n_A\rangle+\langle n_F\rangle},
\label{eq:entro2}
\end{equation}
where $\langle\sigma_{A,F}\rangle$ is the average entropy of an AF or
FM segment and $\langle n_{A,F}\rangle$ is the average number of sites
in a segment.
For reasons discussed above the appropriate way to count boundary sites
between FM and AF segments is to attribute them to the AF side.
Therefore $n_F=\ell-1$ and $n_A=\ell+1$.
The probability to find a FM or AF segment with $\ell$ bonds is
\begin{equation}\begin{array}{ll}
P_F(\ell)=(1-p)p^{\ell-1} &  \hskip 0.5cm \mbox{(FM segments),} \\ & \\
P_A(\ell)=p(1-p)^{\ell-1} &  \hskip 0.5cm \mbox{(AF segments),} \\ &
\end{array} \end{equation}
respectively, and we can calculate the desired averages as
\begin{equation}
\begin{array}{l}
\displaystyle
\langle\sigma_F\rangle=\sum^{\infty}_{\ell=1} P_F(\ell)
\ln[2S(\ell-1)+1], \\
\displaystyle
\langle\sigma_A\rangle=\ln(2S+1)
\sum^{\infty}_{m=1} P_A(2m), \\
\displaystyle
\langle n_F\rangle=\sum^\infty_{\ell=1}P_F(\ell)(\ell-1)
=\frac{p}{1-p}, \\
\displaystyle
\langle n_A\rangle=\sum^\infty_{\ell=1}P_A(\ell)(\ell+1)=\frac{1+p}{p}.
\\
\end{array}
\label{sigma-sigma-n-n}
\end{equation}
In Fig.~\ref{fig:missing} we plot $\Delta\sigma$ as a function of $p$,
which agrees quite well with the estimates of the
missing entropy from the HTE data.

Let us now turn to the Curie behavior of the susceptibility in the
intermediate-temperature regime.
The idea of nearly independent effective spin degrees of freedom
implies, of course, a $1/T$-dependence of the susceptibility as long
as they are uncorrelated.
Analogous to the entropy we can calculate the effective Curie constant
$c$ by averaging the effective spin sizes ($\chi=\mu^2c/k_BT$):
\begin{equation}
c=
\frac{1}{3}
\frac{\langle S_F^2\rangle+\langle S^2_A\rangle}
     {\langle n_F\rangle+\langle n_A\rangle}
\end{equation}
with
\begin{equation} \begin{array}{l}
\displaystyle
\langle S^2_F\rangle=\sum^\infty_{\ell=1}P_F(\ell)S(\ell-1)
[S(\ell-1)+1], \\
\displaystyle
\langle S^2_A\rangle=S(S+1)\sum^\infty_{m=1}P_A(2m), \\
\end{array} \end{equation}
which leads to
\begin{equation}
c=\frac{pS[S(2p^3-4p^2+p-1)+p-1]}{3(1-p)(p-2)}.
\end{equation}
At $p=0.5$ we obtain $c=\frac{1}{8}$ for $S=\frac{1}{2}$ and
$c=\frac{7}{18}$ ($=0.3888$) for $S=1$, both of which agree well with
the values deduced from the HTE and the TMM, $c=0.13$ ($S=\frac{1}{2}$)
and $c=0.4$ ($S=1$).

This discussion of our data is clearly consistent with
the picture of weakly coupled FM and AF segments.
The intermediate temperature regime is aparently well described by an
ensemble of independent AF and FM finite size spin chains with a certain
probability distribution.

\section{Low-energy excitations}
In the previous section we based the interpretation of our data on
the argument that effective spin degrees of freedom are formed
on FM and AF segments which couple only weakly among each other.
In this section we would like to substantiate this picture by discussing
some typical cases and using data obtain from exact diagonalization.
For simplicity we will concentrate on the spin-$\frac{1}{2}$ system.

Let us first consider the case of small $p$ where almost all
bonds are antiferromagnetic.
This system consists of very long AF segments separated by mostly
single FM bonds.
The extreme case of one FM bond between two semi-infinite AF segments
corresponds to a FM impurity bond in an AF spin-$\frac{1}{2}$ chain.
By using the bosonization and the RG techniques, a weak FM coupling can
be shown to be an irrelevant perturbation
with scaling dimension $2$.\cite{Eggert}
Although the FM coupling of strength $J$ in our model is {\it not}
weak, it is very likely that it will be renormalized to zero in the
low-energy limit.\cite{note}
Hence the FM bond surrounded by infinitely long AF segments
will completely decouple the AF segments.
For finite but small $p$ the finite lengths of the AF segments
stops the renormalization of the FM bonds at some energy scale
determined by $\ell\sim 1/p$.
For $p\ll1$ the renormalized FM couplings are so small that we can
neglect them in the first approximation.
A decoupled AF segment of even length form a singlet ground state
while a segment of odd length form a doublet with spin $\frac{1}{2}$.
Thus, in this approximation the degenerate ground state consists of
decoupled AF segments of even and odd length carrying spin $0$ and
$\frac{1}{2}$, respectively.
The residual weak FM couplings will introduce small nearest-neighbor
couplings between the doublets and lift the degeneracy.
Since the even-length segments forms singlets, they do not contribute
to the low-energy degrees of freedom, but they are important to
determine the effective couplings between the spin-$\frac{1}{2}$
segments; two spin-$\frac{1}{2}$ segments separated by an even number
(including $0$) of singlets will couple ferromagnetically while
an odd number of separating singlets makes the effective coupling
antiferromagnetic.
The actual value of the effective coupling depends on the length of
the two spin-$\frac{1}{2}$ segments as well as the number of and
lengths of the separating singlets.
Therefore the resulting effective Hamiltonian is again that of
a random spin-$\frac{1}{2}$ Heisenberg
chain but with random bond strength of either sign.

We have confirmed these properties by exact diagonalization
of finite spin chains for some typical segment configurations.
Figure \ref{fig:antif-1} shows the segmentation of a chain with two
AF segments of odd lengths $5$ and $7$ separated by a single FM bond.
In the spectrum the two lowest energy states, a singlet and a triplet,
are very close in energy ($\Delta E_S=0.13J$), while the gap between
the ground state and the next excited state which involves
intra-segment excitations is considerably higher, $\Delta E_M=0.56J$.
This demonstrates the separation of the low-energy scale from the
intermediate one.
The size of the splitting between the lowest singlet and triplet and
the relative location of these two levels define the magnitude and
sign of the effective coupling:
$J^{\rm eff}=-0.13J$.
In Fig.~\ref{fig:antif-2} we show a chain with two
segments of odd number of bonds separated by one singlet and the
data of their energy levels.
In this case the corresponding energy gaps are $\Delta E_S=0.047J$
and $\Delta E_M=0.65J$, and the separation of energy scales is even
more pronounced.
The relative position of the low-energy singlet and triplet confirms
that the effective coupling in this case is antiferromagnetic.
In Fig.~\ref{fig:antif-3} we link the two configurations in
Figs.~\ref{fig:antif-1} and \ref{fig:antif-2} together.
The spectrum of the effective Hamiltonian,
\begin{equation}
{\cal H}_{\rm eff}=
E_0+\tilde{J}^{\rm eff}_1\tilde{\bf S}_1\cdot\tilde{\bf S}_2
+\tilde{J}^{\rm eff}_2\tilde{\bf S}_2\cdot\tilde{\bf S}_3
\end{equation}
with $\tilde{J}^{\rm eff}_1=0.047J$ and $\tilde{J}^{\rm eff}_2=-0.125J$
estimated from Figs.~\ref{fig:antif-1} and \ref{fig:antif-2}
(dashed lines), agrees very well with the lowest part of the exact
spectrum (solid lines), showing that the low-energy degrees of freedom
are well described by a nearest-neighbor Heisenberg Hamiltonian.
The energy shift $E_0=-7.65J$ is adjusted so that the ground-state
energies agree.
The fact that the true spectrum is reproduced so well with effective
coupling calculated from two-segment chains confirms that possible
non-nearest-neighbor interactions in the effective Hamiltonian
are small.

Let us now turn to the case of $p$ close to 1, the nearly FM chain.
Although different in structure, we can argue quite similarly to
$p\ll1$ on the separation of the low-energy scale.
Consider two FM segments of finite length separated by a single
AF bond.
This AF bond has a tendency to lock its two adjacent spins into
a singlet, which effectively decouples the two FM segments.
In a modified chain where all AF bonds are much stronger than the FM
ones, this tendency is even more pronounced, and in the limit of
infinite AF exchange the FM segments are completely decoupled.
In this limit each FM segment forms a local ground state of maximum
spin, and the overall ground state is highly degenerate since the
large segment-spins are non-interacting.
A finite $J_{\rm AF}$ lifts this degeneracy, and the effective
Hamiltonian for the segment-spins is, to first order in
$J_{\rm FM}/J_{\rm AF}$, an AF nearest-neighbor Heisenberg Hamiltonian
with coupling strengths depending on the lengths of the FM segments.
Our exact diagonalization results suggest that this picture remains
true even when $J_{\rm FM}/J_{\rm AF}=1$ and that this
results in a separation of energy scales.
In Fig.~\ref{fig:ferro-1} is a FM chain with a single AF bond.
{}From the segmentation we expect five almost degenerate multiplets of
spin $S=\frac{1}{2},\frac{3}{2},\ldots,\frac{9}{2}$ corresponding
to different relative orientations between the effective spins
$\tilde{S}_1=2$ and $\tilde{S}_2=\frac{5}{2}$.
These multiplets indeed appear in the spectrum
[Fig.~\ref{fig:ferro-1}(b)] and are reproduced by the effective
Hamiltonian
\begin{equation}
{\cal H}=E_0
+\sum^4_{n=1}\tilde{J}_{n}(\tilde{\bf S}_1\cdot\tilde{\bf S}_2)^n
\label{eq:ferroeff-1}
\end{equation}
with $E_0=-2.72J$, $\tilde{J}_1=8.5\times10^{-3}J$,
$\tilde{J}_2=3.2\times10^{-4}J$, $\tilde{J}_3=2.0\times10^{-5}J$,
and $\tilde{J}_4=-7.8\times10^{-7}J$.
In the spectrum in Fig.~\ref{fig:ferro-1}(b) we can also clearly
identify the four multiplets $S=\frac{1}{2},\ldots,\frac{7}{2}$
corresponding to the right segment forming a
spin-$\frac{3}{2}$ state (magnon excitation) and the three multiplets
$S=\frac{3}{2},\frac{5}{2},$ and $\frac{7}{2}$ corresponding to
the left segment forming a spin-1 state.
The excitation energies to these one-magnon states are
$\Delta E_{M_1}=0.22J$ and $\Delta E_{M_2}=0.31J$, respectively.
Finally we form a three-segment chain in Fig.~\ref{fig:ferro-2}.
The lower part of the true spectrum is once again in excellent
agreement with the spectrum of the effective Hamiltonian which is
taken to be
\begin{equation}
{\cal H}_{\rm eff}=
E_0+\tilde{J}_1^{\rm eff}\tilde{\bf S}_1\cdot\tilde{\bf S}_2
+\tilde{J}_2^{\rm eff}\tilde{\bf S}_2\cdot\tilde{\bf S}_3.
\label{eq:ferroeff-2}
\end{equation}
In Eq.~(\ref{eq:ferroeff-2}) $\tilde{J}_2^{\rm eff}=0.0085J$ is
estimated from Fig.~\ref{fig:ferro-1} and
$\tilde{J}_1^{\rm eff}=0.011J$ is obtained from a similar calculation.
The energy shift is adjusted to $E_0=-4.28J$ for a best fit.
These results confirm that the low-energy physics for spin chains with
$p$ close to one is also well described by a random nearest-neighbor
Heisenberg Hamiltonian.
However, in contrast to small $p$ the effective couplings for $p$
close to one are all antiferromagnetic, though the magnitude of the spins
are random.

For intermediate values of $p$ the situation is more complex
because the typical configuration of FM and AF bonds leads to
a sequence of rather short segments.
The extreme case is a sequence of alternating FM and AF bonds which is
known to favor a singlet (dimer) configuration with an excitation gap.
Such a sequence, located between two segments with finite effective spin
degrees of freedom, yields extremely small couplings.
On the other hand, we can also find the situation that a FM segment
borders on an AF segment with an even number of bonds, which
forms an effective spin doublet.
In this case the effective coupling between the two segments is not
necessarily small.
Nevertheless, the coupling is small enough to form the separate segment
spins which in turn govern the low-energy physics.
This type of configurations would, of course, be the first to undergo
intersegment correlation as the temperature is lowered.
Hence, we expect that in the effective low-energy Heisenberg model
the distribution of the coupling strengths as well as spin sizes is
rather broad for these systems.

\section{Conclusions}

We have studied the 1D quantum Heisenberg model with random FM and
AF bond disorder by means of several numerical methods.
A comparison of the quantum spin chains with the corresponding
classical system reveals interesting similarities.
At first sight the differences seem to be only of quantitative nature.
However, through a careful analysis of our data, we found that the
low-temperature (low-energy) physics of the quantum and classical
system has profound differences.
While the quantum spin system exhibits a separation of two energy
scales between the low and intermediate energies, only one scale
is present in the classical case.

The idea of the creation of weakly-interacting effective spin
degrees of freedom in FM and AF segments along the chain plays
a key role in understanding the physics of the quantum system.
While the high-energy (high-temperature) behavior is determined
by the individual spins of the chain, they are replaced by new effective
spin degrees of freedom towards the low-energy limit.
As we have demonstrated in the last section, the low-energy
physics is well captured by an effective model which has the structure
of a nearest-neighbor Heisenberg chain with spins of variable size and
FM and AF interactions of random strength.
As a consequence three temperature regimes emerge in our quantum spin
system (in contrast to two in the classical case):
the high-, intermediate- and low-temperature regime.
The latter two are clearly determined by the effective
low-energy model.
In this sense the intermediate temperature regime may be considered
as the high-temperature regime of the effective model.
Therefore the effective spins yield a Curie-like susceptibility in
this regime as confirmed in Section IIIC.
In a similar way the specific heat may be separated into two
contributions.
One is due to the individual spins, which generate a specific heat peak
when they start to correlate at a temperature $k_BT\sim J$.
The other originates from the effective spins, which give rise to
another presumably much weaker peak-like structure containing,
however, a considerable fraction of the entropy at lower temperature.

The methods we used in this study did not allow us to investigate
the low-energy physics extensively. Alternative methods have to be
applied here.
The investigation of the thermodynamics and magnetic properties of this
regime is in progress and will be presented elsewhere.

\acknowledgments
We are grateful to H.-C.\ zur Loye and T.\ Nguyen for stimulating
discussion on their experimental results.
Furthermore, we would like to thank T.K.\ Ng, V.L.\ Pokrovsky, and
W.\ Putikka for helpful discussions.
K.B.T.\ and N.N.\ express their gratitude to
K.\ Kubo for his help on the program of the transfer matrix calculation.
We are also grateful for the financial support by Swiss Nationalfonds
(M.S., No.\ 8220-037229), by the Swedish Natural Science Research
Council (E.W.), and by the NSF through the MRSEC program under Grant
No.\ DMR-94-00334.

\end{multicols}

\narrowtext

\begin{figure}
\caption{Specific heat of the spin-$\frac{1}{2}$ chain for
$p=0$ (solid), 0.25 (dotted), 0.5 (long dashed),
0.75 (short dashed), and 1 (dashed dotted)
obtained from two-point Pad\'e approximation.
The filled circles denote the results by Bl\"ote for the uniform chains
and the empty circles the data of our transfer matrix calculation.}
\label{fig:c-s=1/2}
\end{figure}

\begin{figure}
\caption{Specific heat of the spin-1 chain for $p=0$ (solid),
0.25 (dotted), 0.5 (long dashed), 0.75 (short dashed), and 1 (dashed dotted)
obtained from Dlog Pad\'e approximation.
The filled circles denote the results by Bl\"ote for the uniform chains
and the empty circles the data of our transfer matrix calculation.}
\label{fig:c-s=1}
\end{figure}

\begin{figure}
\caption{Specific heat divided by temperature for the spin-$\frac{1}{2}$
chain.
The notations are the same as in Fig.\ \protect\ref{fig:c-s=1/2}.}
\label{fig:c/T-1/2}
\end{figure}

\begin{figure}
\caption{Specific heat divided by temperature for the spin-1 chain.
The notations are the same as in Fig.\ \protect\ref{fig:c-s=1}.}
\label{fig:c/T-1}
\end{figure}

\begin{figure}
\caption{Inverse susceptibility of the spin-$\frac{1}{2}$ chain
for $p=$ 0, 0.25, 0.5, 0.75 and 1.}
\label{fig:chi-1/2}
\end{figure}

\begin{figure}
\caption{Inverse susceptibility of the spin-$\frac{1}{2}$ and spin-1
chains at $p=0.5$.}
\label{fig:chi-1}
\end{figure}

\begin{figure}
\caption{Missing entropy.  The solid ($S=\frac{1}{2}$) and dashed
($S=1$) curves are calculated from Eqs.\ (\protect\ref{eq:entro2})
and (\protect\ref{sigma-sigma-n-n}).
The circles (diamonds) are the missing entropies estimated from
the HTE for $S=\frac{1}{2}$ ($S=1$).}
\label{fig:missing}
\end{figure}

\begin{figure}
\caption{(a) The segmentation of a short chain containing one FM bond.
AF couplings are represented with $+$ and FM couplings with $-$.
(b) The energy spectrum of the chain in (a).}
\label{fig:antif-1}
\end{figure}

\begin{figure}
\caption{(a) The segmentation of a chain with two FM bonds resulting
in two odd-length segments separated by one singlet.
Bonds are represented as in Fig.\ \protect\ref{fig:antif-1}.
(b) The energy spectrum of the chain in (a).}
\label{fig:antif-2}
\end{figure}

\begin{figure}
\caption{(a) The segmentation of a chain of length $L=19$.
The segments appearing in this particular chain are the ones in
Figs.\ \protect\ref{fig:antif-1} and \protect\ref{fig:antif-2}.
Bonds are represented as in Fig.\ \protect\ref{fig:antif-1}.
(b) The energy spectrum of the chain in (a) (soild lines)
and the spectrum of the correspinding effective Hamiltonian for
the segment-spins (dashed lines).}
\label{fig:antif-3}
\end{figure}

\begin{figure}
\caption{(a) An AF bond in a FM chain locks its two adjecent spins
into a singlet.
This effectively decouples the FM segments which form local
ground states of largest possible spins.
Bonds are represented as in Fig.\ \protect\ref{fig:antif-1}.
(b) The spectrum of the chain in (a).}
\label{fig:ferro-1}
\end{figure}

\begin{figure}
\caption{(a) The segmentation of a nearly FM chain resulting in
three weakly interacting large segment-spins.
Bonds are represented as in Fig.\ \protect\ref{fig:antif-1}.
(b) The energy spectrum of the chain in (a) (solid lines)
and the spectrum of the corresponding effective Hamiltonian
(dashed lines).}
\label{fig:ferro-2}
\end{figure}

\clearpage

\mediumtext
\begin{table}
\caption{Coefficients $u(l,m)$ for the spin-$\frac{1}{2}$ chain;
 $u(K;p)=(J/4)\sum_l\sum_mu(l,m)(\beta J/4)^lp^m$.}
\label{table:u(l,m)s=1/2}
\begin{tabular}{cdcdcd}
$(l,m)$ & $u(l,m)$ &  $(l,m)$ & $u(l,m)$  & $(l,m)$ & $u(l,m)$ \\
\hline
(1,0) & $-$3.0           &(12,3)& 7713.500413     &(18,3)& $-$1134121.9573 \\
(2,0) & $-$3.0           &(13,0)& 8196.990854     &(18,4)& $-$288649.3274  \\
(2,1) & 6.0              &(13,1)& $-$25934.104588 &(18,5)& 115459.7310     \\
(3,0) & 5.0              &(13,2)& 23117.786328    &(19,0)& $-$2079.9826    \\
(4,0) & 15.0             &(13,3)& 5632.636519     &(19,1)& $-$646413.7595  \\
(4,1) & $-$30.0          &(13,4)& $-$2816.318258  &(19,2)& 2645440.1027    \\
(5,0) & $-$4.2           &(14,0)& 6660.909634     &(19,3)& $-$3976953.5968 \\
(5,1) & $-$20.8          &(14,1)& $-$39516.948771 &(19,4)& 1935729.0738    \\
(5,2) & 20.8             &(14,2)& 78429.276975    &(19,5)& 63297.2706      \\
(6,0) & $-$61.13333333   &(14,3)& $-$51765.812859 &(19,6)& $-$21099.0874   \\
(6,1) & 122.26666667     &(14,4)& $-$780.557689   &(20,0)& $-$1251808.4999 \\
(7,0) & $-$40.48571429   &(14,5)& 312.223089      &(20,1)& 5117166.1965    \\
(7,1) & 201.75238095     &(15,0)& $-$28013.413242 &(20,2)& $-$7236338.9668 \\
(7,2) & $-$201.75238095  &(15,1)& 80562.547491    &(20,3)& 2810586.2334    \\
(8,0) & 204.90476190     &(15,2)& $-$47628.744538 &(20,4)& 3017931.4549    \\
(8,1) & $-$381.46666667  &(15,3)& $-$65867.605907 &(20,5)& $-$1202116.2591 \\
(8,2) & $-$85.02857143   &(15,4)& 32933.802930    &(20,6)& $-$5056.3158    \\
(8,3) & 56.68571428      &(16,0)& $-$54542.64412  &(20,7)& 1444.6371       \\
(9,0) & 353.8973545      &(16,1)& 255075.95809    &(21,0)& $-$1184631.917  \\
(9,1) & $-$1276.5291005  &(16,2)& $-$433951.49101 &(21,1)& 7782735.876     \\
(9,2) & 1276.5291005     &(16,3)& 275899.26561    &(21,2)& $-$19645416.929 \\
(10,0)& $-$500.5733333   &(16,4)& 20102.59259     &(21,3)& 23366818.910    \\
(10,1)& 571.1106878      &(16,5)& $-$8041.03709   &(21,4)& $-$10787051.462 \\
(10,2)& 1290.1079365     &(17,0)& 66386.55357     &(21,5)& $-$1075629.595  \\
(10,3)& $-$860.0719576   &(17,1)& $-$123625.66748 &(21,6)& 358543.182      \\
(11,0)& $-$1918.0408850  &(17,2)& $-$161124.99951 &(22,0)& 4217217.366     \\
(11,1)& 6375.6162771     &(17,3)& 568820.19930    &(22,1)& $-$15639609.441 \\
(11,2)& $-$6238.2222222  &(17,4)& $-$282707.26277 &(22,2)& 16514166.308    \\
(11,3)& $-$274.7881097   &(17,5)& $-$2043.40448   &(22,3)& 5968948.429     \\
(11,4)& 137.3940551      &(17,6)& 681.13439       &(22,4)& $-$25376121.949 \\
(12,0)& 259.304916       &(18,0)& 293845.3269     &(22,5)& 9967514.773     \\
(12,1)& 3338.140375      &(18,1)& $-$1270211.3634 &(22,6)& 182933.980      \\
(12,2)& $-$11570.250620  &(18,2)& 1989832.2634    &(22,7)& $-$52266.733    \\
\end{tabular}
\end{table}

\clearpage

\begin{table}
\caption{Coefficients $\chi(l,m)$ for the spin-$\frac{1}{2}$ chain;
 $\chi(K;p)=\beta\mu^2\sum_l\sum_m\chi(l,m)(\beta J/4)^lp^m$.}
\label{table:chi(l,m)s=1/2}
\begin{tabular}{cdcdcd}
$(l,m)$  & $\chi(l,m)$ & $(l,m)$ & $\chi(l,m)$ & $(l,m)$ &  $\chi(l,m)$ \\
\hline
(0,0)& 0.25             &(6,6)& 32.0             &(9,9) & 256.0           \\
(1,0)& $-$0.5           &(7,0)& 1.015873016      &(10,0)& $-$59.9102293   \\
(1,1)& 1.0              &(7,1)& 8.679365080      &(10,1)& 24.7207055      \\
(2,1)& $-$2.0           &(7,2)& $-$0.133333333   &(10,2)& $-$134.8949559  \\
(2,2)& 2.0              &(7,3)& $-$74.577777778  &(10,3)& $-$241.7721340  \\
(3,0)& 0.666666667      &(7,5)& 224.0            &(10,4)& 508.1876543     \\
(3,1)& 0.666666667      &(7,6)& $-$224.0         &(10,5)& 1173.6380953    \\
(3,2)& $-$6.0           &(7,7)& 64.0             &(10,6)& $-$1585.8793651 \\
(3,3)& 4.0              &(8,0)& 18.12857143      &(10,7)& $-$2048.0       \\
(4,0)& 0.833333333      &(8,1)& $-$7.32063492    &(10,8)& 4352.0          \\
(4,1)& 2.0              &(8,2)& 57.50476191      &(10,9)& $-$2560.0       \\
(4,2)& 6.0              &(8,3)& 38.29841270      &(10,10)& 512.0          \\
(4,3)& $-$16.0          &(8,4)& $-$237.81587302  &(11,0)& $-$109.9150746  \\
(4,4)& 8.0              &(8,5)& $-$96.0          &(11,1)& 300.2769921     \\
(5,0)& $-$1.4           &(8,6)& 629.33333333     &(11,2)& 70.4828219      \\
(5,1)& $-$1.866666667   &(8,7)& $-$512.0         &(11,3)& $-$338.9203527  \\
(5,2)& 6.0              &(8,8)& 128.0            &(11,4)& $-$1239.6134039 \\
(5,3)& 22.666666667     &(9,0)& 13.1816578       &(11,5)& 983.6675837     \\
(5,4)& $-$40.0          &(9,1)& $-$52.9643739    &(11,6)& 4324.9777778    \\
(5,5)& 16.0             &(9,2)& $-$23.9428571    &(11,7)& $-$3283.7079367 \\
(6,0)& $-$4.433333333   &(9,3)& 180.4486772      &(11,8)& $-$6912.0       \\
(6,1)& $-$0.888888889   &(9,4)& 259.9365079      &(11,9)& 10922.6666667   \\
(6,2)& $-$20.444444444  &(9,5)& $-$656.5079365   &(11,10)& $-$5632.0      \\
(6,3)& 10.666666667     &(9,6)& $-$522.6666667   &(11,11)& 1024.0         \\
(6,4)& 74.666666667     &(9,7)& 1685.3333333     & \\
(6,5)& $-$96.0          &(9,8)& $-$1152.0        & \\
\end{tabular}
\end{table}

\begin{table}
\caption{Coefficients $u(l,m)$ for the spin-1 chain;
 $u(K;p)=J\sum_l\sum_mu(l,m)(\beta J/2)^lp^m$.}
\label{table:u(l,m)s=1}
\begin{tabular}{cdcdcd}
$(l,m)$ & $u(l,m)$ & $(l,m)$ & $u(l,m)$ & $(l,m)$ & $u(l,m)$ \\ \hline
(1,0)& $-$2.6666666667  &(10,2)& 28.8792214       &(15,1)& 45953.430459    \\
(2,0)& $-$1.3333333333  &(10,3)& $-$19.2528142    &(15,2)& $-$45757.817154 \\
(2,1)& 2.6666666667     &(11,0)& 481.3638086      &(15,3)& $-$391.226630   \\
(3,0)& 5.9259259259     &(11,1)& 1301.8758468     &(15,4)& 195.613379      \\
(4,0)& 6.6666666667     &(11,2)& $-$1301.3134946  &(16,0)& 54519.198734    \\
(4,1)& $-$13.333333333  &(11,3)& $-$1.1247047     &(16,1)& $-$99299.036124 \\
(5,0)& $-$18.251851852  &(11,4)& 0.5623535        &(16,2)& $-$29214.214665 \\
(5,1)& $-$1.896296296   &(12,0)& 2977.0623249     &(16,3)& 19463.245209    \\
(5,2)& 1.896296296      &(12,1)& $-$5835.6140989  &(16,4)& 19.346878       \\
(6,0)& $-$31.664197531  &(12,2)& $-$355.5316528   &(16,5)& $-$7.738804     \\
(6,1)& 63.328395062     &(12,3)& 237.0211016      &(17,0)& 28441.68270     \\
(7,0)& 58.441544189     &(13,0)& $-$849.9549193   &(17,1)& $-$249359.16806 \\
(7,1)& 23.237154614     &(13,1)& $-$8019.3116116  &(17,2)& 246954.37836    \\
(7,2)& $-$23.237154614  &(13,2)& 8006.2448768     &(17,3)& 4809.45489      \\
(8,0)& 147.08759553     &(13,3)& 26.1334724       &(17,4)& $-$2404.41627   \\
(8,1)& $-$293.62210464  &(13,4)& $-$13.0667454    &(17,5)& $-$0.37317      \\
(8,2)& $-$1.65925926    &(14,0)& $-$12916.439014  &(17,6)& 0.12400         \\
(8,3)& 1.10617283       &(14,1)& 24667.952605     &(18,0)& 1688755.7953    \\
(9,0)& $-$179.33842257  &(14,2)& 3494.640784      &(18,1)& $-$5177380.6581 \\
(9,1)& $-$190.19539704  &(14,3)& $-$2329.308912   &(18,2)& 5503050.1273    \\
(9,2)& 190.19539704     &(14,4)& $-$0.677419      &(18,3)& $-$4013509.8333 \\
(10,0)& $-$669.2886525  &(14,5)& 0.270973         &(18,4)& 517214.6225     \\
(10,1)& 1328.9508980    &(15,0)& $-$1530.190525   &(18,5)& $-$206885.8485  \\
\end{tabular}
\end{table}

\clearpage

\begin{table}
\caption{Coefficients $\chi(l,m)$ for the spin-1 chain;
 $\chi(K;p)=\beta\mu^2\sum_l\sum_m\chi(l,m)(\beta J/2)^lp^m$.}
\label{table:chi(l,m)s=1}
\begin{tabular}{cdcdcd}
$(l,m)$ & $\chi(l,m)$ & $(l,m)$ & $\chi(l,m)$ & $(l,m)$ & $\chi(l,m)$ \\
\hline
(0,0)& 0.6666666667    &(5,3)& 299.19341564    &(8,0)& $-$5.234080     \\
(1,0)& $-$1.7777777778 &(5,4)& $-$449.49245542 &(8,1)& 6.320244        \\
(1,1)& 3.5555555556    &(5,5)& 179.79698217    &(8,2)& 425.727623      \\
(2,0)& 1.4814814815    &(6,0)& 4.14229538      &(8,3)& 1053.738742     \\
(2,1)& $-$9.4814814815 &(6,1)& $-$8.95473251   &(8,4)& $-$1911.972048  \\
(2,2)& 9.4814814815    &(6,2)& $-$148.3676269  &(8,5)& $-$7884.430625  \\
(3,0)& 1.580246914     &(6,3)& $-$164.8139003  &(8,6)& 18539.066606    \\
(3,1)& 9.481481481     &(6,4)& 1281.0534979    &(8,7)& $-$13637.934055 \\
(3,2)& $-$37.925925926 &(6,5)& $-$1438.3758573 &(8,8)& 3409.483514     \\
(3,3)& $-$25.283950617 &(6,6)& 479.4586191     &(9,0)& $-$67.132623    \\
(4,0)& $-$2.13991770   &(7,0)& 21.0825005      &(9,1)& $-$461.172347   \\
(4,1)& 6.32098765      &(7,1)& 50.8237507      &(9,2)& 2085.379854     \\
(4,2)& 61.10288066     &(7,2)& 85.6220698      &(9,3)& $-$10591.213175 \\
(4,3)& $-$134.84773663 &(7,3)& $-$633.0976376  &(9,4)& 42515.683875    \\
(4,4)& 67.42386831     &(7,4)& $-$1373.4491693 &(9,5)& $-$74434.761484 \\
(5,0)& $-$5.37283951   &(7,5)& 5024.3267795    &(9,6)& 57428.487934    \\
(5,1)& $-$19.45459534  &(7,6)& $-$4474.9471117 &(9,7)& $-$16408.139410 \\
(5,2)& 0.70233196      &(7,7)& 1278.5563176    & \\
\end{tabular}
\end{table}

\widetext
\begin{table}
\caption{The ground state energy estimated from Dlog Pad\'e approximants.}
\label{table:energy}
\begin{tabular}{cddddddddddd}
$p$ & 0 & 0.1 & 0.2 & 0.3 & 0.4 & 0.5 & 0.6 & 0.7 & 0.8 & 0.9 & 1 \\
\hline
$S=1/2$ & $-$0.439 & $-$0.428 & $-$0.414 & $-$0.398 & $-$0.380 &
$-$0.360 & $-$0.337 & $-$0.316 & $-$0.293 & $-$0.273 & $-$0.240 \\
$S=1$ & $-$1.395\tablenotemark[1] & $-$1.35 & $-$1.308 & $-$1.267 &
$-$1.222 & $-$1.183 & $-$1.143 & $-$1.104 & $-$1.06 & $-$1.04 & $-$1.01 \\
\end{tabular}
\tablenotetext[1]{This value is in reasonable agreement with the result
obtained by Nightingale and Bl\"ote:\cite{Nightingale}
$E_0=-1.4015J$.}
\end{table}


\begin{references}
\bibitem[*]{AF}
 On leave from Dept.\ of Applied Physics, University of Tokyo,
 Hongo, Bunkyo-ku, Tokyo 113, Japan.
\bibitem[\dag]{KT} Present address: Department of Physics,
 University of California at Los Angels, Los Angels, CA 90024.
\bibitem{Bulaevskii} L.N.\ Bulaevskii, A.V.\ Zvarykina, Yu.S.\ Karimov,
 R.B.\ Lyubovskii, and I.F.\ Shchegolev,
 Sov.\ Phys.\ JETP {\bf 35}, 384 (1972).
\bibitem{Hirsch} J.E.\ Hirsch and J.V.\ Jos\'e, Phys.\ Rev.\ B {\bf 22},
 5339 (1980); J.E.\ Hirsch, {\it ibid.}\ {\bf 22}, 5355 (1980).
\bibitem{Ma} S.-k.\ Ma, C.\ Dasgupta, and C.K.\ Hu,
 Phys.\ Rev.\ Lett.\ {\bf 43}, 1434 (1979);
 C.\ Dasgupta and S.-k.\ Ma, Phys.\ Rev.\ B {\bf 22}, 1305 (1980).
\bibitem{Fisher} D.S.\ Fisher, Phys.\ Rev.\ B {\bf 50}, 3799 (1994).
\bibitem{Bhatt} R.N.\ Bhatt and P.A.\ Lee, Phys.\ Rev.\ Lett.\ {\bf 48},
 344 (1982).
\bibitem{Nagaosa} N.\ Nagaosa, J.\ Phys.\ Soc.\ Jpn.\ {\bf 56}, 2460
 (1987); Phys.\ Rev.\ B {\bf 39}, 2188 (1989) and references therein.
\bibitem{Doty} C.A.\ Doty and D.S.\ Fisher, Phys.\ Rev.\ B {\bf 45},
 2167 (1992).
\bibitem{Haas} S.\ Haas, J.\ Riera, and E.\ Dagotto, Phys.\ Rev.\ B
 {\bf 48}, 13174 (1993).
\bibitem{Runge} K.J.\ Runge and G.T.\ Zimanyi, Phys.\ Rev.\ B {\bf 49},
 15212 (1994).
\bibitem{Wilkinson} A.P.\ Wilkinson, A.K.\ Cheetham, W.\ Kunnman, and
 A.\ Kvick, Eur.\ J.\ Solid State Inorg.\ Chem.\ {\bf 28}, 453 (1991).
\bibitem{Nguyen} T.\ N.\ Nguyen, Ph.~D thesis, Massachusetts Institute of
 Technology, 1994 (unpublished); T.\ N.\ Nguyen, D.\ M.\ Giaquinta,
 and H.-C.\ zur Loye, Chem.\ Mater.\ {\bf 6}, 1642 (1994).
\bibitem{Gelfand} For detailed explanation of high-temperature series,
 see M.\ P.\ Gelfand, R.R.P.\ Singh, and D.A.\ Huse,
 J.\ Stat.\ Phys.\ {\bf 59}, 1093 (1990);
 see also D.F.B.\ ten Haaf and J.M.J.\ van Leeuwen,
 Phys.\ Rev.\ B {\bf 46}, 6313 (1992).
\bibitem{Rushbrooke} G.S.\ Rushbrooke, J.\ Math.\ Phys.\ {\bf 3},
 1106 (1964).
\bibitem{Baker} G.A.\ Baker Jr., G.S.\ Rushbrooke, and H.E.\ Gilbert,
 Phys.\ Rev.\ {\bf 135}, A1272 (1964).
\bibitem{Guttmann} A.J.\ Guttmann,
 {\it Phase Transitions and Critical Phenomena}, Vol.\ 13, ed.\ by
 C.\ Domb and J.L.\ Lebowitz (Academic Press, London, 1989);
 D.S.\ Gaunt and A.J.\ Guttmann, {\it Phase Transitions and Critical
 Phenomena}, Vol.\ 3, ed.\ by C.\ Domb and M.S.\ Green
 (Academic Press, London, 1974).
\bibitem{Honda} N.\ Honda and H.\ Igarashi, J.\ Phys.\ Soc.\ Jpn.\ {\bf 53},
 2930 (1984).
\bibitem{differential} In addition to the Pad\'e approximation,
 we tried the method of differential approximants\cite{Guttmann}
 to analyze the high-temperature series.
 However, we found that the differntial approximants were rather unstable,
 and we could not get better results than the Pad\'e approximation.
\bibitem{Suzuki1} M.\ Suzuki, Phys.\ Rev.\ B {\bf 31}, 2957 (1985);
 H.\ Betsuyaku, Prog.\ Theor.\ Phys.\ {\bf 73}, 319 (1985);
 S.\ Takada and K.\ Kubo, J.\ Phys.\ Soc.\ Jpn.\ {\bf 55}, 1671 (1986).
\bibitem{Suzuki2} M.\ Suzuki, Phys.\ Lett.\ {\bf 113A}, 299 (1985).
\bibitem{Blote} H.W.J.\ Bl\"ote, Physica {\bf 79B}, 427 (1975).
\bibitem{Furusaki} A.\ Furusaki, M.\ Sigrist, P.A.\ Lee, K.\ Tanaka,
 and N.\ Nagaosa, Phys.\ Rev.\ Lett.\ {\bf 73}, 2622 (1994).
\bibitem{Reed} A similar result is obtained by P.\ Reed,
 J.\ Phys.\ A {\bf 25}, 5861 (1992).
\bibitem{Tonegawa} T.\ Tonegawa, H.\ Shiba, and P.\ Pincus,
 Phys.\ Rev.\ B {\bf 11}, 4683 (1975); T.\ Tonegawa, {\it ibid.}\ {\bf 14},
 3166 (1976).
\bibitem{Fahnle} M.\ F\"ahnle, Phys.\ Stat.\ Sol.\ (b){\bf 100},
 K113 (1980).
\bibitem{MEFisher} M.E.\ Fisher, Am.\ J.\ Phys.\ {\bf 32}, 343 (1964).
\bibitem{Takahashi} M.\ Yamada and M.\ Takahashi,
 J.\ Phys.\ Soc.\ Jpn.\ {\bf 55}, 2024 (1986);
 M.\ Takahashi, Phys.\ Rev.\ Lett.\ {\bf 58}, 168 (1987).
\bibitem{Eggert} S.\ Eggert and I.\ Affleck, Phys.\ Rev.\ B {\bf 46},
 10866 (1992).
\bibitem{note} In the limit of strong FM coupling,
 the two spins connected by the FM bond form a triplet, and the system
 is reduced to the Kondo problem:\cite{Eggert} an $S=1$ spin is coupled
 antiferromagnetically to two semi-infinite $S=\frac{1}{2}$
 AF Heisenberg chains.  In the low-energy limit, the $S=1$
 spin and two $S=\frac{1}{2}$ spins form a Kondo singlet.
 Therefore the limit of the strong FM coupling is not a stable
 fixed point.  If there is no nontrivial intermediate-coupling fixed
 point, the FM coupling will always be renormalized to zero.
\bibitem{Nightingale} M.P.\ Nightingale and H.W.J.\ Bl\"ote,
 Phys.\ Rev.\ B {\bf 33}, 659 (1986).
\end{references}
\end{document}